\documentclass[twocolumn,english,superscriptaddress,prl,amsmath,amssymb,floatfix,longbibliography,nofootinbib]{revtex4-2}
\usepackage[T1]{fontenc}
\usepackage[utf8]{inputenc}
\usepackage{color}
\usepackage{babel}
\usepackage{amsmath}
\usepackage{graphicx}
\usepackage{esint}
\usepackage[unicode=true,pdfusetitle,
 bookmarks=true,bookmarksnumbered=false,bookmarksopen=false,
 breaklinks=false,pdfborder={0 0 0},pdfborderstyle={},backref=false,colorlinks=true]
 {hyperref}

\makeatletter
\usepackage{dcolumn}
\usepackage{bm}
\usepackage{color}
\usepackage{soul}
\usepackage{babel}
\usepackage{amsfonts}
\usepackage{slashed}
\usepackage{enumerate}
\usepackage{subfigure}
\usepackage{mathtools}
\usepackage{babel}

\makeatother

\DeclareMathOperator{\Tr}{Tr}
\newcommand{\bra}[1]{\left\langle #1 \right|}
\newcommand{\ket}[1]{\left| #1 \right\rangle}
\newcommand{\braket}[2]{\bra{#1} #2\rangle}

\begin{document}

\title{Super-resolution enhancement in bi-photon spatial mode demultiplexing}
\author{Florence Grenapin}
\affiliation{Nexus for Quantum Technologies, University of Ottawa, Ottawa, K1N 6N5, ON, Canada}
\author{Dilip Paneru}
\affiliation{Nexus for Quantum Technologies, University of Ottawa, Ottawa, K1N 6N5, ON, Canada}
\author{Alessio D'Errico}
\affiliation{Nexus for Quantum Technologies, University of Ottawa, Ottawa, K1N 6N5, ON, Canada}
\author{Vincenzo Grillo}
\affiliation{Istituto Nanoscienze, Consiglio Nazionale delle Ricerche, Via G. Campi 213/A, 41125 Modena, Italy}
\author{Gerd Leuchs}
\affiliation{Nexus for Quantum Technologies, University of Ottawa, Ottawa, K1N 6N5, ON, Canada}
\affiliation{Max Planck Institute for the Science of Light, 91058 Erlangen, Germany}
\author{Ebrahim Karimi}
\affiliation{Nexus for Quantum Technologies, University of Ottawa, Ottawa, K1N 6N5, ON, Canada}
\affiliation{Max Planck Institute for the Science of Light, 91058 Erlangen, Germany}

\begin{abstract}
Imaging systems measuring intensity in the far field succumb to Rayleigh's curse, a resolution limitation dictated by the finite aperture of the optical system. Many proof-of-principle and some two-dimensional imaging experiments have shown that, by using spatial mode demultiplexing (SPADE), the field information collected is maximal, and thus, the resolution increases beyond the Rayleigh criterion. Hitherto, the SPADE approaches are based on resolving the lateral splitting of a Gaussian wavefunction. Here, we consider the case in which the light field originates from a bi-photon source, i.e. spontaneous parametric down-conversion, and a horizontal separation is introduced in one of the two photons. We show that a separation induced in the signal photon arm can be super-resolved using coincidence measurements after projecting both photons on Hermite-Gauss modes. Remarkably the Fisher information associated with the measurement is enhanced compared to the ordinary SPADE techniques by $\sqrt{K}$, where $K$ is the Schmidt number of the two-photon state that quantifies the amount of spatial entanglement between the two photons.
\end{abstract}
\maketitle

\section*{Introduction}
The resolution of far-field optical imaging systems based on direct intensity measurements is limited by the Point Spread Function (PSF), a diffraction phenomenon dictated by light wavelength and aperture width of the optics involved~\cite{LordRayleigh}. The accuracy of this measurement diverges for small separations of two-point sources; an effect also referred to as ``Rayleigh's curse''. The discovery of this phenomenon has encouraged research efforts in other fields of imaging, such as near field imaging~\cite{Scan,NearField} or imaging based on electronic effects~\cite{SEM, STM}, which goes beyond the optical Rayleigh limit. In 2016, Tsang and his colleagues proposed a simple spatial mode demultiplexing (SPADE) scheme, which is robust to the resolution curse~\cite{OGTsang}. It shows that our ability to estimate a separation between two incoherent point sources collapses as the separation approaches zero when using conventional intensity measurements, thus assimilating Rayleigh's curse to an intrinsic loss of Fisher information at small separations. The SPADE scheme proposed uses spatial mode sorting, with mode projections tailored to the PSF of the imaging system, which collects maximal Fisher information regardless of the separation magnitude. This has prompted many further theoretical and experimental findings in recent years, exploring applications and modifications of SPADE for different imaging systems~\cite{ArbitraryPSF,ArbitraryPSF_Luis,French,Steinberg}, coherent point sources~\cite{Coherent1,Coherent2,Coherent3}, limitations in the presence of cross-talk or other noise sources~\cite{FrenchCrossTalk,Noise,santamaria2022balanced}, and even extensions to higher-dimensional objects~\cite{MultiParameter,2DFI}. Although the explicit model estimating all points of an arbitrary two-dimensional (2D) object is vastly complicated to derive in terms of Fisher information, 2D imaging simulations~\cite{Kaden} and experiments~\cite{Oxford} using post-processing algorithms (deconvolution, Machine Learning) have shown an increase in resolution beyond the Rayleigh limit. This solidifies the intuition relating the distinguishability of two-point sources to the overall resolution of a 2D imaging system.\newline
The main idea behind SPADE is to estimate the separation between two-point sources by looking at the alteration of the field's phase, in particular, at the change in phase symmetry, which can be efficiently probed by the Hermite-Gauss (HG) decomposition for the usual Gaussian PSF. In this work, we address the following question: \textit{Can the spatial correlations emerging from two-photon sources enhance the resolution sensitivity?} We show that it is possible to super-resolve a separation using mode-sorting in coincidence measurements with a PSF that is itself entangled in the form of SPDC light. More specifically, we consider a two-photon wavefunction which allows a Schmidt decomposition in HG modes, with Schmidt number $K$. An incoherent displacement, which mimics the effect of a transmitting sample, is applied on one of the two photons, and the displacement is estimated by projecting the resulting state on direct products of HG modes. We note that the somewhat related inverse problem of generating a spatially or temporally shaped pure photon state in one beam of an entangled pair (ghost interference) likewise requires the projection on pure states in the other beam of the entangled pair \cite{sych2017}. Our main result shows that the Fisher information, whose inverse gives a lower bound on the estimation error, scales as $\sqrt{K}/2$, where $K=1$ corresponds to the separable case of a Gaussian point spread function. Hence any spatially entangled two-photon source provides an advantage with respect to the ordinary SPADE. Here, we derive this result and present a first proof-of-principle experimental implementation. We will conclude by discussing the possible settings in which the two-photon SPADE can give a practical advantage.

\section*{Theory}
To understand the resolution limits of a given experimental technique, one must evaluate the lower bound on the achievable uncertainty of the estimator $\delta$ (which, in our case, is the lateral displacement). This limit is given by the Cramer-Rao bound~\cite{OGTsang, QMetrology}: the estimator's standard error is lower bounded by the inverse of the Fisher information $\Delta \delta \geq 1/\sqrt{n\,\mathcal{FI}}$~\cite{giovannetti2011advances,helstrom1969quantum}, where $n$ corresponds to the number of repeated measurements and the Fisher information, $\mathcal{FI}$, can be calculated from the probabilities $\mathcal{P}_j$ of a given measurement outcome $j$ as $\mathcal{FI}= \sum_{j}\frac{1}{\mathcal{P}_j}\left(\frac{\partial \mathcal{P}_j}{\partial \delta}\right)^2$. In the following, we consider the problem of resolving an incoherent transverse displacement of a photon which is in a spatially correlated state with another idler photon. We will calculate the Fisher information for a bi-photon spatial mode demultiplexing scheme and compare the result with the classical SPADE, which emerges from the uncorrelated limit of our scenario.

In a typical imaging setup, the image plane of a point source is described by a two-dimensional PSF $\Psi(\mathbf{r})$ (with $\mathbf{r}=(x,y)$). In the case of two incoherent point sources separated horizontally by a distance $s$, the wavefunction in the image plane can be described by the mixed state: $\rho = \frac{1}{2}\left(\ket{\Psi^+}\bra{\Psi^+}+\ket{\Psi^-}\bra{\Psi^-}\right)$, where $\braket{\text{x}}{\Psi^{\pm}}:= \Psi(x \pm \frac{s}{2},y)$. An intensity measurement, which corresponds to a projective measurement in the position bases, results in the outcomes with a distribution, $I(x,y)= \frac{1}{2}\left(|\Psi^+|^2+|\Psi^-|^2 \right)$. The Fisher information for the direct imaging method is known to rapidly fall to zero for separations smaller than the width of the point spread function~\cite{OGTsang}. However, by mode sorting using modes that form orthogonal bases for the space containing the PSF (SPADE), the total FI remains constant across all values of $s$~\cite{OGTsang}. Optimal bases are ones in which most of the FI for small $s$ is captured in the first couple of measurements, making schemes like binary SPADE~\cite{Steinberg} simple and attractive. We show in the Supplementary Material~\ref{SM:optimal projection} how a projection on the derivative of the PSF with respect to the coordinate associated with the direction of the displacement is an optimal projection for the small separation regime. More rigorous ways to determine optimal measurement bases have been developed~\cite{ArbitraryPSF}.

In a coincidence imaging setup with entangled SPDC photon pairs, there is no pure wavefunction describing one of the single photons' quantum states. The single particle state is maximally mixed, which can mimic a Gaussian in its intensity pattern when viewed on the camera. The pure state describing the entangled particles is a bi-photon state of signal and idler which can be expressed in the Schmidt basis of Hermite-Gauss (HG) modes~\cite{SpatialSchmidt}: 
%%%
\begin{equation}\label{eq:spdc}
    \ket{\Psi}=\sum_{m,n}C_{mn}\ket{m,n}_s\otimes\ket{m,n}_i,
\end{equation}
%%%
where the $C_{mn}$ coefficients are the Schmidt coefficients of the HG decomposition and $\ket{m,n}$ states are the 2D HG modes of order $(m,n)$ with a beam waist parameter $\sigma_s$: $\braket{x,y}{m,n}:=\text{HG}_{m,n}(x,y)=\mathcal{N}\exp(-(x^2+y^2)/\sigma_s^2)\,\text{H}_m(\sqrt{2} x/\sigma_s)\,\text{H}_n(\sqrt{2} y/\sigma_s)$, where $\text{H}_m(x)$ are Hermite polynomials and $\mathcal{N}$ a normalization constant. In this basis, the correlations are diagonal: a signal photon in a particular $\ket{i,j}$ mode is entangled with an idler photon in that exact mode (see the Supplementary Material \ref{SM: schmidt decomposition}). In practice, any action occurring on one photon, such as the typical diffraction causing PSFs, is also reflected in the bi-photon wavefunction. The Schmidt coefficients can be given an analytical expression (see \cite{fedorov2009gaussian,miatto2012cartesian}): $C_{m,n}=|(\gamma+1)/(\gamma+1)|^{m+n}4\gamma/(1+\gamma)^2$, where $\gamma = \frac{1}{\sigma_p}\sqrt{\frac{L\lambda_p}{2\pi}}$ is a parameter defined by crystal length $L$, pump beam waist $\sigma_p$ and wavelength $\lambda_p$. Here, we investigate how two-photon SPADE allows us to super-resolve separation introduced in one of the photon paths as if the PSF was of the form of the source (just as in \cite{Steinberg,French}) -- which is the bi-photon wavefunction described above. \\

Finding the optimal basis of projection for SPADE is not as straightforward as in the single photon PSF cases. Although the derivative of the bi-photon PSF with respect to the signal photon's coordinates can be calculated directly (Supplementary Material \ref{SM: schmidt decomposition}), this is a nonseparable state for which it is challenging to devise a single-shot projector experimentally. However, we note that interesting progress is being made in that direction~\cite{thekkadath2018projecting}). Instead, as in ordinary SPADE, we consider the projection on the basis of direct products of HG modes $\{ \ket{m,n}_{s}\!{\bra{m,n}} \otimes \ket{m,n}_{i}\!{\bra{m,n}} \}$. The expression for a projection onto mode $\text{HG}_{k,l}(x_i,y_i)$ in idler and $\text{HG}_{k',l'}(x_s,y_s)$ in signal can be derived as,
\begin{equation} \label{eq:coincprob}
    \prescript{k',l'}{k,l}{\mathcal{P}_{coinc}} = \frac{1}{2} C_{k',l}^2\delta_{l,l'}  \sum_{\pm} |\braket{k}{k'^{\pm}}|^2.
\end{equation}
With $s =0$ the correlations are diagonal: only projections where $k=k'$ have non-zero outcomes. As $s$ increases, contributions from the first off-diagonal modes appear. Equation~\eqref{eq:coincprob} can be given an analytical expression using the result: $\braket{m}{n^{\pm}} = \sqrt{\frac{m!}{n!}}2^{\frac{m-n}{2}}(\pm s)^{n-m}e^{-\frac{s^2}{4}}\mathcal{L}_m^{n-m}(\frac{s^2}{2})$, where $\mathcal{L}_i^{\alpha}$ is the generalized Laguerre polynomial of order $i,\,\alpha$, $n\geq m$ (see Supplementary Material \ref{SM: overlap calculation} for the derivation).

For small values of $s$, we consider the expansion of $\prescript{k',l'}{k,l}{\mathcal{P}_{coinc}}$ to the order of $s^2$  by substituting $\ket{m^{\pm}} = \ket{m} \pm \frac{s}{2}\ket{m}' + O(\frac{s^2}{4})$ into Eq.~\eqref{eq:coincprob}. The result $\mathcal{P}_{kl}^{k'l'} \approx C_{k',l}^2 \Big[ \delta_{k,k'}\big \{1-\frac{s^2}{2}(2k'+1)\big \} + \delta_{k,k'-1}\frac{k'}{2} +\delta_{k,k'+1}\frac{k'+1}{2} \Big]\delta_{l,l'}$ implies that for each mode $k$ considered in the signal beam, only the neighbouring $k'$ modes give nonzero coincidence counts: $\mathcal{P}_{kl}^{kl} \approx C_{k,l}^2 (1-\frac{s^2}{2}(2k+1))$, $\mathcal{P}_{kl}^{k+1,l} \approx s^2C_{k+1,l}^2\frac{k+1}{2}$, $\mathcal{P}_{kl}^{k-1,l} \approx s^2C_{k-1,l}^2\frac{k}{2}$. Of these terms, only the $k=k', k'\pm1$ modes varying quadratically with the separation will translate to a nonzero FI in the small separation regime, $s\rightarrow 0$ (see Supplementary Material \ref{SM:HG_biphotonSPADE}):
%%%%%%
\begin{equation} \label{eq:FIspade}
    \begin{cases}      \mathcal{FI}_{kl}^{kl}(s) \approx \frac{\big[C_{k,l}^2(2k+1)s\big]^2}{4-2s^2(2k+1)} \rightarrow 0\\ \\
      \prescript{k+1}{k}{\mathcal{FI}_{coinc}} = \frac{1}{2}(k+1)C_{k+1,l}^2\\ \\
      \prescript{k-1}{k}{\mathcal{FI}_{coinc}} = \frac{1}{2}kC_{k-1,l}^2
    \end{cases}.
\end{equation}
%%%%%%
Since the FI in these particular projections is independent of the separation $s$, the variance of our separation estimates does not suffer from the so-called Rayleigh's curse, similar to the SPADE technique for a single photon.\newline
%%%%%
\begin{figure}[t]
  \centering
    \includegraphics[scale=0.6]{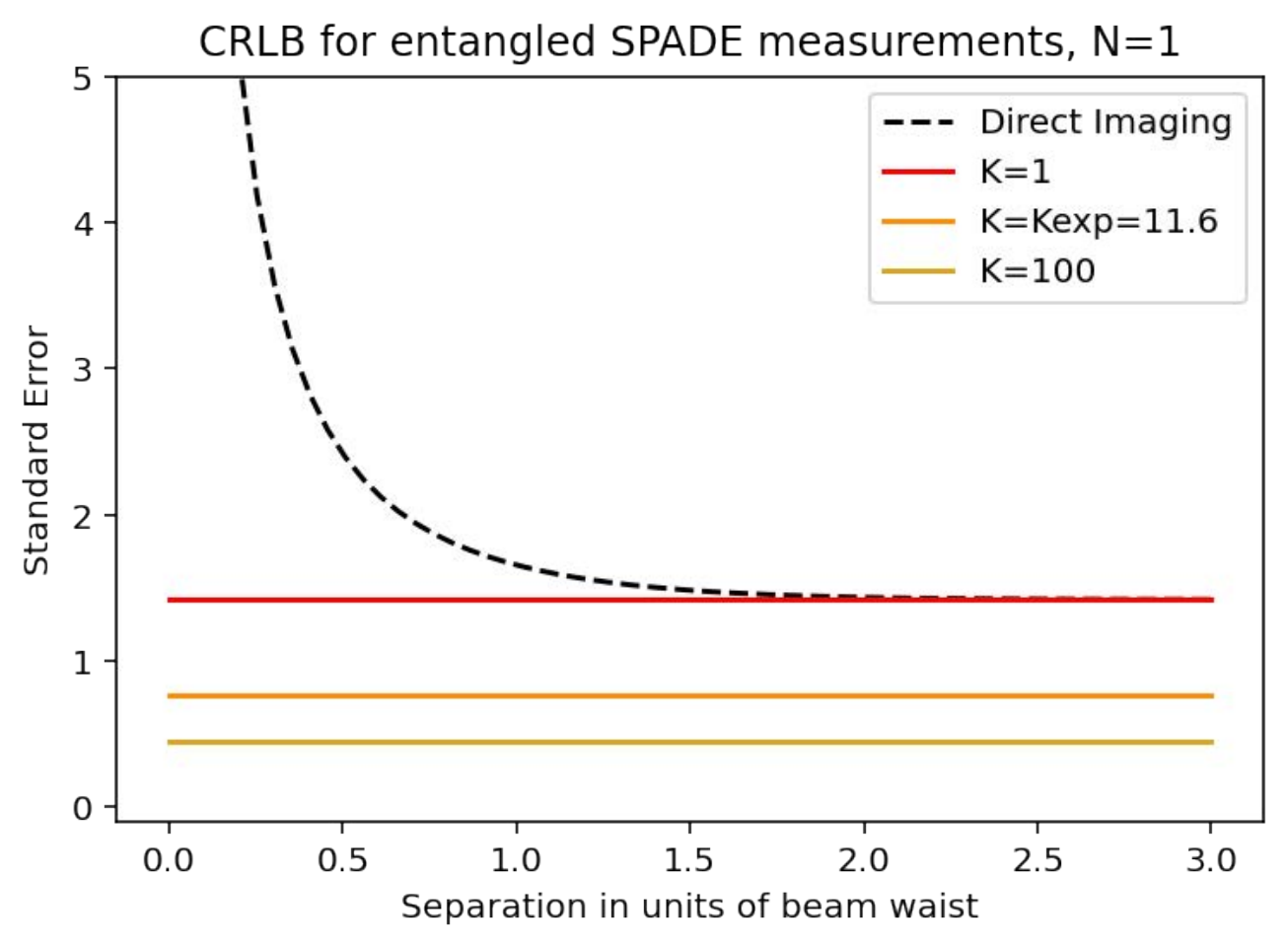}
    \caption{Classical SPADE measurements (lower bounded by $K=1$) attain the SQL precision. Higher values of entanglement provide an advantage over classical SPADE.}
  \label{fig:CRLBtheoretical}
\end{figure}
%%%%%

The total Fisher Information can then be obtained by summing the individual contributions from all of the detected modes. Interestingly, for small values of $s$, the total has an upper bound dictated by the strength of the spatial correlation between the photon pair. More precisely, 
\begin{equation}
      \prescript{tot}{2D}{\mathcal{FI}_{coinc}} = \gamma +\gamma^{-1} = \frac{\sqrt{K}}{2} \geq \frac{1}{2},
\end{equation}
where $K=\frac{1}{4}(\gamma+\frac{1}{\gamma})^2$ is the Schmidt rank often used to quantify the strength of bipartite entanglement (see Supplementary Material \ref{SM: schmidt decomposition} for the derivation). This leads to a corresponding lower bound in the achievable precision as,
\begin{equation}
    \Delta \hat{s}^2 \geq \frac{2}{\sqrt{K}},
\end{equation}
where $\Delta \hat{s}^2$ is the variance in the separation estimates. The result suggests that the precision in the measurement of near-zero separations is limited only by the strength of the spatial correlations, thereby suggesting high dimensional entanglement as a promising resource for super-resolution measurements. Not only does the method circumvent the so-called ``Rayleigh curse'' like the ordinary SPADE (which is bounded by the special case of  $K=1$, see Fig.~\ref{fig:CRLBtheoretical}), but it also offers further enhancement proportionate to the correlations in the system. In what follows, we describe in detail an experimental implementation and results of the method outlined above, along with some simulation results.

\section*{METHODS}
For the purpose of a simpler experimental implementation in the laboratory, we considered a smaller subspace spanned by modes with $l=0$, (i.e.  $\text{HG}_{00},\text{HG}_{01},\text{HG}_{02},\ldots,\text{HG}_{0n}$) for both the signal and idler photons. The total Fisher information associated with this set of joint projective measurements can be written as (see Supplementary Material \ref{SM:HG_biphotonSPADE}),
\begin{equation} \label{eq:totalFIspade}
    \prescript{tot}{1D}{\mathcal{FI}_{coinc}}= \frac{1}{2} + \frac{1}{2}\bigg(\frac{1-\gamma}{1+\gamma}\bigg)^2.
\end{equation}
Experimentally, we reconstructed the correlation matrices in the Hermite Gauss (HG) basis, for a bi-photon state generated via spontaneous parametric downconversion (SPDC), for different traverse separations. The separation was then inferred from the observed correlation matrices using  Maximum Likelihood Estimation.\newline
\begin{figure*}[htp]
  \centering
  \subfigure[Experimental Setup]{\includegraphics[scale=0.22]{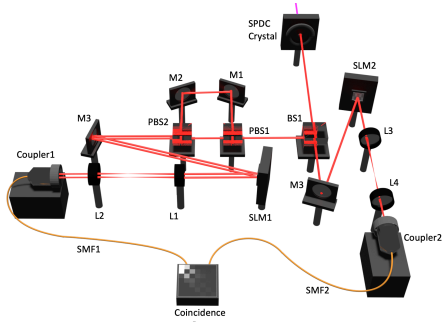}}\quad
  \subfigure[Theoretical (top) and Experimental (bottom) correlation matrices]{\includegraphics[scale=0.18]{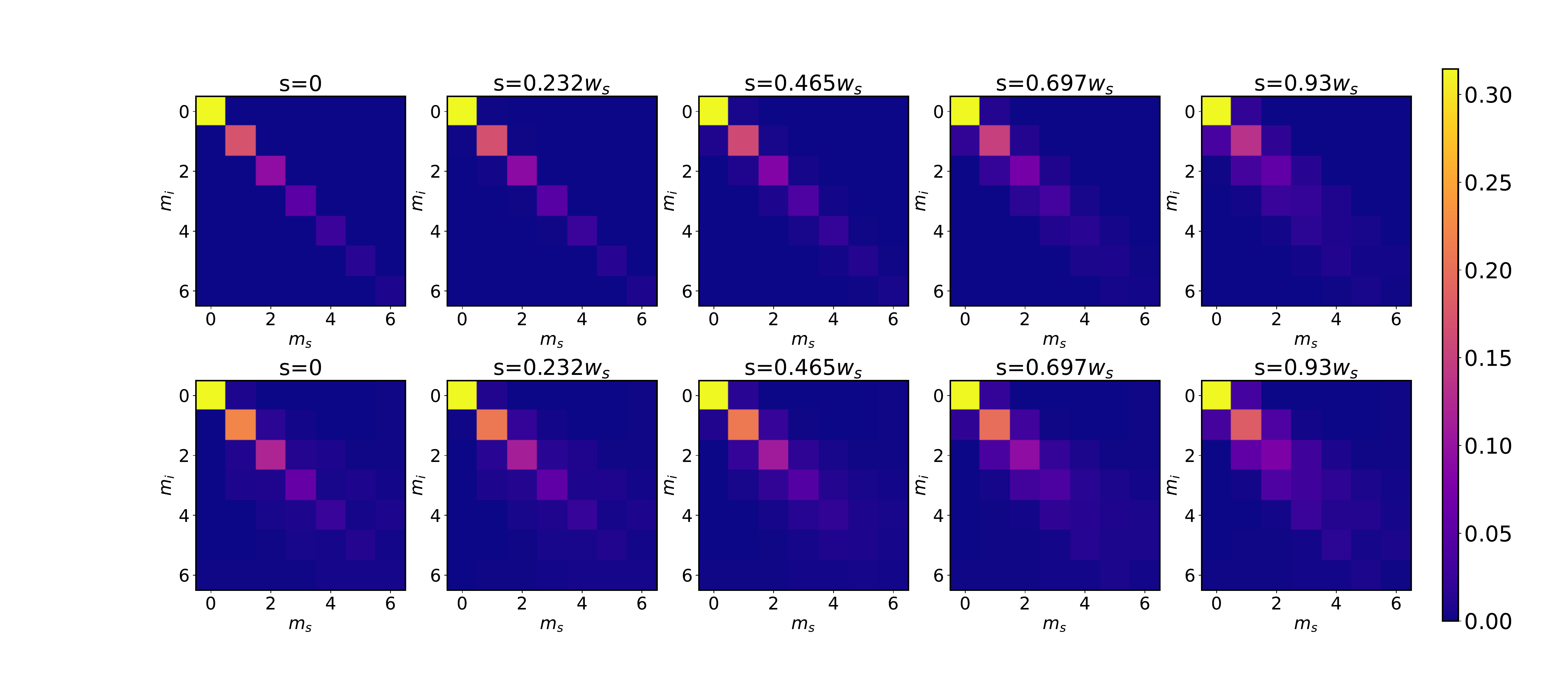}}
  \caption{a) A 405 nm laser shines a 1-mm type I BBO crystal generating degenerate downconverted photon pairs. Idler and signal photon are separated by a 50:50 beam splitter (BS). An unbalanced MZI interferometer implements an incoherent lateral displacement on the signal photon. The two photons are then projected on HG modes by coupling to single mode fibers (SMF) after being spatially modified by holograms displayed on spatial light modulators (SLM1 and SLM2). The photons are then detected by a pair of single photon avalanche diode (SPAD) detectors then the coincidence events are detected via a coincidence counting module. Legend: BS = Beam Splitter; PBS = Polarizing Beam Splitter; M = Mirror; SLM = Spatial Light Modulator; L= Lens; SMF = Single Mode Fiber. b)Theoretical (top) and Experimental (bottom) cross-talk matrices showing coincidence counts for first the 7x7 joint projections on signal and idler photons, for 5 different values of separation {$s$ between 0 and $0.93\sigma_s$.} We can see the off-diagonal coefficients  becoming more prominent as the separation increases.}
  \label{Setup}
\end{figure*}

\noindent\textbf{Experimental Realization:} The downconverted photon pairs are generated by pumping a 0.5-mm-thick type-I Barium Borate crystal (BBO) with a 200-mW 405-nm pump beam obtained from the Second Harmonic Generation (SHG) of a 1-GHz 100-fs laser beam. The pump beam is focused to a waist of $\sigma_p=40~\mu$m at the BBO crystal, which sets a value of $\gamma=0.15$. The value of $\gamma$ determines the waist of the Schmidt modes, the Schmidt coefficients in the Hermite Gauss basis, and the Schmidt rank $K$~\cite{miatto2012cartesian,straupe2011angular}. The generated photon pairs are then split into two paths by a 50:50 beamsplitter (BS), and the signal photon travels through an asymmetrical Mach-Zender Interferometer (MZI), which introduces a transverse separation between the Horizontal (H), and the Vertical (V) polarization components of the beam. Two mirrors are placed on translation stages, specific combinations of which allow us to generate a tunable symmetric separation about the centre of the undisplaced beam. The MZI also ensures that a path difference of $\approx $ 20cm exists between the horizontal (H) and the vertical (V) polarization components, which ensures that we have two incoherent point sources. In each of the signal and the idler arms, projections onto the spatial Schmidt modes are realized via the amplitude flattening technique~\cite{Ebrahim, Fred}, whereby a combination of appropriate holograms displayed in Spatial Light Modulators (SLM) along with suitable demagnification to the fibre plane approximates the ideal projective measurement for a particular spatial mode. The first order of the diffracted light is selected with a pinhole and coupled into a single mode fibre (SMF) after being filtered by bandpass filters ($810\pm 5$)~nm to ensure the collection of degenerate pairs. Each SMF is then connected to a Single Photon Avalanche Diode (SPAD) detectors (Excelitas SPCM-AQRH-14-FC), and coincidence counts are registered via a time-correlated single photon counting system. 
Further details on the experimental setup with the essential components are illustrated in Fig.~\ref{Setup}. \newline

To determine the experimental Schmidt waist, multiple sets of data for differing waists of projection were taken for a zero displacement, and the experimental projection waist was chosen so as to have as close to a diagonal density matrix as possible. For each applied separation, the setup was realigned so that the centre of the displaced beam was coupled to the fibre. The coincidence counts in the diagonal modes for zero separation account for at least 82\% of the total counts in the $7\times7$ cross-talk matrix. The deviation from a perfectly diagonal decomposition can be attributed to several factors, including a deviation of the pump beam shape from a Gaussian mode and modal cross-talk in the detection scheme. These factors can be tackled, in principle, by improved control of the pump shape (e.g. through spatial filtering or the use of SLMs) and the use of other detection schemes, respectively. Indeed, the recent advances in spatial mode sorting through multi-plane converters allow, in principle, a lossless and low cross-talk detection scheme~\cite{fontaine2019laguerre}.

\subsection*{Results}
We applied evenly spaced lateral displacements at increments of $0.0465\,\sigma_s$, from $s=0$ to $s=1.35\,\sigma_s$, and recorded the outcomes of the joint measurements onto modes $\text{HG}_{n_i,m_i}$ in the idler arm and $\text{HG}_{n_s,m_s}$ in the signal arm for $n_{i/s} =0$, and $m_{i/s} = 0,1, \ldots ,6$. Each set of data is then a $7\times7$ matrix containing 49 joint projection outcomes, the coincidence counts for each of which were accumulated for up to 60s. The $7\times7$ matrix sums were normalized to account for laser intensity fluctuations. For such small separations, to a very good approximation, it can be safely assumed that almost all of the generated photons are in the $7\times7$ Hilbert space. In total, for each separation, $\approx 37,000$ photons were collected. The results for a few sets of displacements, along with the corresponding theoretical predicted outcomes (Eq.~\eqref{eq:coincprob}), are illustrated in the form of correlation matrices (see Fig. \ref{Setup}b)). The figure shows that the experimental outcomes closely match the corresponding theoretical predictions. We also see the modes just outside the diagonal starting to increase in significance symmetrically with the increasing separation.
\begin{figure}[h]
  \includegraphics[scale=0.55]{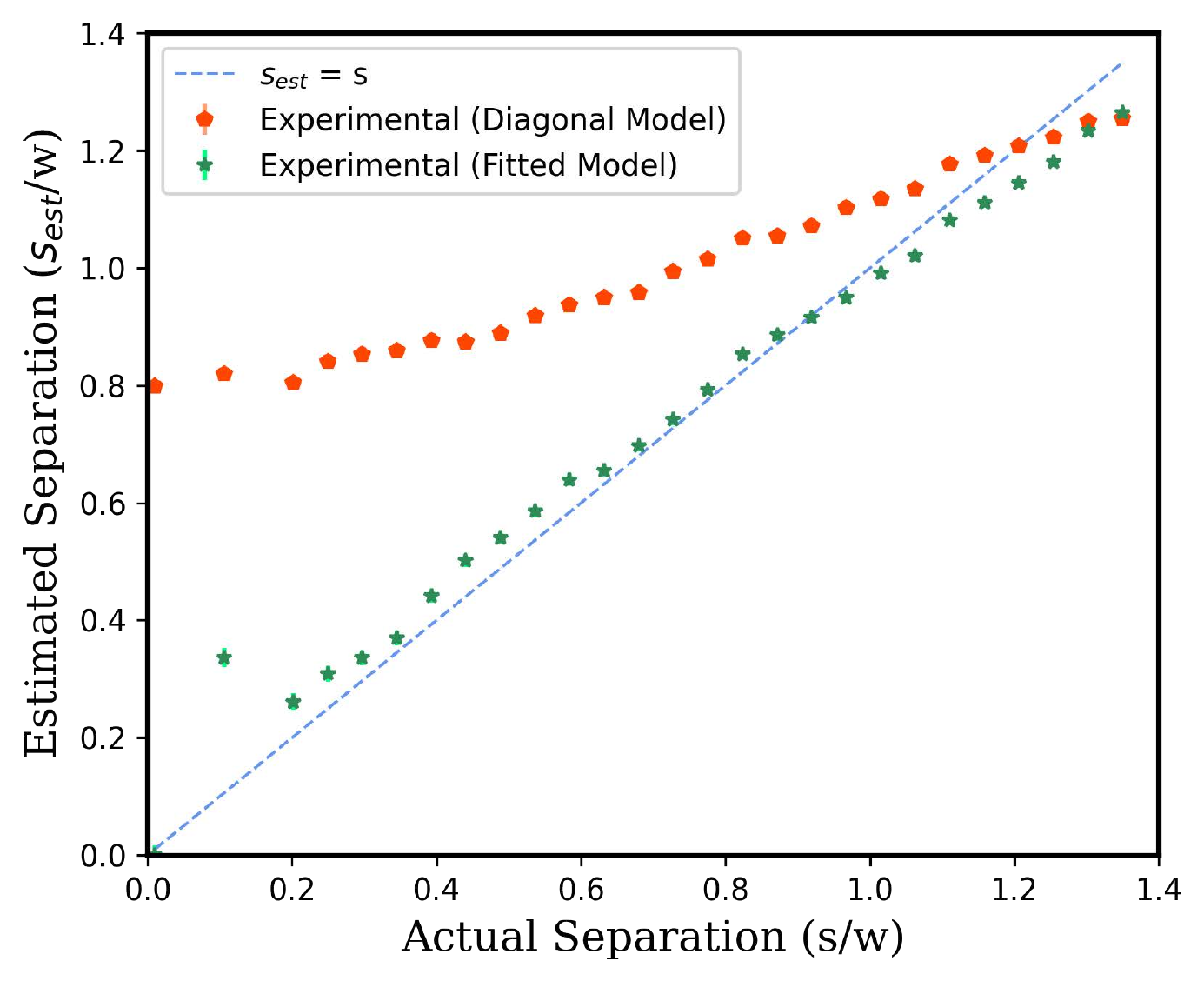}
  \caption{Estimates from the data vs  the Actual Separation. The estimation was first performed assuming an ideal theoretical model, i.e. a perfectly diagonal mode decomposition (red) where probabilities are defined by Eqn.~\eqref{eq:coincprob}. We see a significant deviation from the ideal $s_{est}=s$ estimates, most likely due to modal cross-talk, attenuation, and dark counts. Therefore, in order to account for these effects, a linear fitting of the counts, as functions of the theoretical probabilities, was performed before the estimation, the results of which closely match the ideal curve (green).}
  \label{fig:estvsactual}
\end{figure}
\begin{figure}[ht]
  \centering
   {\includegraphics[scale=0.55]{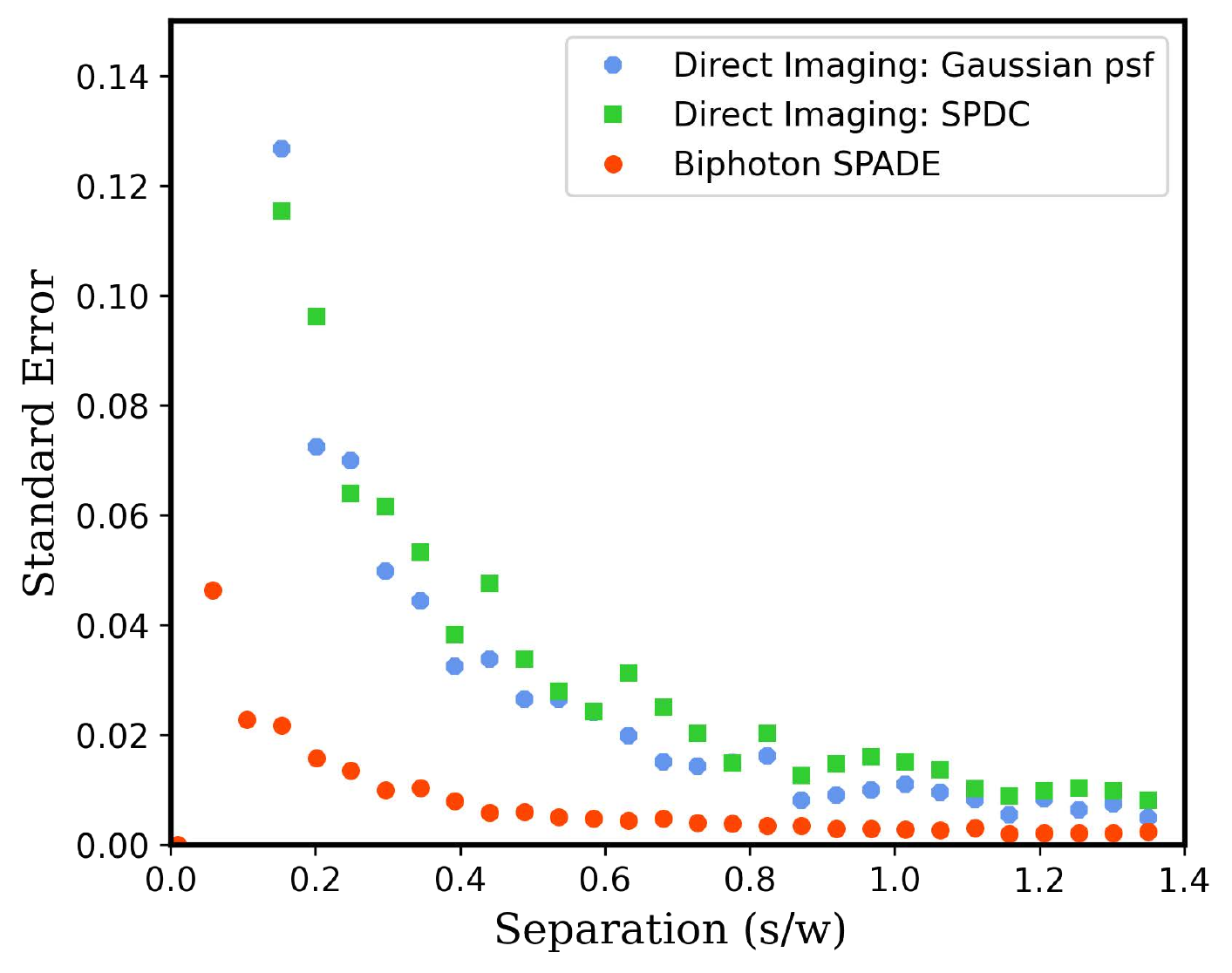}}\quad
   \caption{Simulation results for Biphoton SPADE compared with Direct Imaging of both Gaussian and SPDC PSFs for N = 37,000 photons. A 50-pixel array was used for direct imaging simulation, corresponding to intensity measurements at 50 different points. Similarly, for the biphoton SPADE ($K= 11.6$), the simulation was done for 49 mode projections with combinations of Hermite Gauss modes with indices $n =0, m=0,\ldots,6$ for each signal and idler. A maximum likelihood estimation was performed on generated samples, and the standard error in the separation estimates was plotted.}
  \label{fig:Sims}
\end{figure}
In order to obtain the estimates of the separation parameter $s$, first, we perform the least squares fitting for the normalized counts (or equivalently probabilities) to the expression. $P_{ij}= \alpha_{ij} \mathcal{P}_{ij} + \beta_{ij}$, where $\mathcal{P}_{ij}$s are the theoretical probabilities given by Eq.~\eqref{eq:coincprob}, $\alpha_{ij}$ is a scaling factor to account for attenuation for different modes, and an additive factor $\beta_{ij}$ to account for the spurious photon counts.  For each separation, we then applied a Maximum Likelihood Estimation procedure (MLE), maximizing the likelihood function defined as $L(s):= (N!/\prod{n_{ij}!})\prod_{ij} {P}_{ij}(s)^{n_{ij}}$,  where $n_{i,j}$'s are the photon counts for the joint projection onto  $\text{HG}_{0i}$ for signal and $\text{HG}_{0j}$ for the idler, and $N$ is the total photon counts in all of the modes for that particular separation. The results of the estimation are shown in Fig.~(\ref{fig:estvsactual}).\newline

In order to compare the precision in the estimates obtained from this method with direct imaging, we also performed simulation runs of the experimental outcomes for the direct imaging of the SPDC point sources whose mode decomposition is characterized by $\gamma = 0.15$. The simulation results are plotted in Fig~(\ref{fig:Sims}). We see that the biphoton SPADE outperforms the direct imaging of the SPDC and also the direct imaging of the Gaussian with the same waist as the Schmidt's waist, especially offering an order of magnitude advantage for near zero separation.

\section*{Discussion and Conclusion}
We have theoretically derived optimal projective measurements for separation estimation using a biphoton wavefunction generated via the SPDC process. The Fisher information for projection onto these modes remains constant even for near zero separation, which shows that the measurement scheme is not plagued by the so-called ``Rayleigh's curse'' inherent in Direct Imaging. Moreover, the precision in the proposed method is  limited only by the strength of the spatial correlations inherent in the photon pairs. In particular, we show that the total Fisher information of the separation estimation is proportional to the square root of the Schmidt rank ($K$) of the system. Therefore, depending upon the strength of the correlations in the system, the method even offers further enhancement that scales with $K$ compared to ordinary SPADE, which corresponds to the  trivial case of $K=1$. As a proof of concept, we also experimentally implemented the scheme using down-converted photon pairs and obtained results that are in good agreement with the theoretical predictions. We also performed simulations comparing the precision in our estimates with the case of Direct Imaging, where we indeed observe at least an order of magnitude improvements, especially for near zero separations, where the errors start to diverge. Our theoretical and experimental results suggest the potential of using high-dimensional entanglement for further enhancing super-resolution measurements. Combined with neural networks (see, e.g. Ref. \cite{Oxford}), this technique can provide a new tool for high-resolution imaging in low-light conditions of passive samples with variable transmissivity.
As a perspective, we believe that these ideas may have an appeal also in electron microscopy where the resolution problem is very relevant especially when trying to separate two close point-like objects like atoms. Optimal measurements (to which SPADE belongs) in the single electron framework have been already studied \cite{troiani2020efficient} but the recent advances in controlling entangled electrons \cite{meier2022few, haindl2022coulomb} could pave the way to increase the resolution below 10 pm, which would be the highest spatial resolution achieved so far. More generally, the Schmidt decomposition of electron pair could provide a new framework for advanced metrology in quantum transmission electron microscopy.
\newline
\noindent\textbf{Acknowledgement:} E.K. acknowledges the scientific discussions with Professor Aephraim Steinberg. This work was supported by the Canada Research Chair (CRC) Program, NRC-uOttawa Joint Centre for Extreme Quantum Photonics (JCEP) via the Quantum Sensors Challenge Program at the National Research Council of Canada, and the European Union’s Horizon 2020 Research and Innovation Programme under grant agreement No 766970 (QSORT).

\bibliography{paper}
\clearpage
\onecolumngrid
\renewcommand{\figurename}{\textbf{Figure}}
\setcounter{figure}{0} \renewcommand{\thefigure}{\textbf{S{\arabic{figure}}}}
\setcounter{table}{0} \renewcommand{\thetable}{S\arabic{table}}
\setcounter{section}{0} \renewcommand{\thesection}{Section S\arabic{section}}
\setcounter{equation}{0} \renewcommand{\theequation}{S\arabic{equation}}
\onecolumngrid

\begin{center}
{\Large Supplementary Information for: \\Super-resolution enhancement in bi-photon spatial mode demultiplexing}
\end{center}
%\appendix
\vspace{1 EM}

\section{\label{SM:optimal projection} Optimal Projection}

\noindent Fisher information describes the amount of information a measurement contains about a certain parameter being estimated. The Cramer-Rao Lower Bound, which is the minimum possible variance on the unbiased estimation of the parameter, is given by the inverse of the Fisher Information\cite{giovannetti2011advances},
\begin{equation} \label{eq:crlbSM}
\sigma_{crlb} \geq \frac{1} {\sqrt{\mathcal{FI}}} \geq 
\frac{1}{\sqrt{Tr[\frac{\partial \rho}{\partial \delta}L]}} 
\end{equation}
where $\rho$ is the state of the system, $\delta$ is the parameter to be estimated and $L$ is the so called symmetric logarithmetic operator defined as 
$
\rho L +L\rho = 2\frac{\partial \rho}{\partial \delta}.
$
\noindent Higher the Fisher Information associated with a particular set of measurement, better the precision in the estimates.\\ In the super resolution schemes for separation estimation with the so called Spatial Mode Demultixplexing (SPADE), a projective measurement is carried out in spatial modes (most commonly in the Hermite Gauss basis). The Fisher Information of a parameter $\delta$, in the basis of such set of spatial modes $\{ \ket{j}\bra{j} \}$  is given by,

\begin{equation} \label{eq:fisherSM}
\mathcal{FI} = \sum_{j}\frac{1}{\mathcal{P}_j}\left(\frac{\partial \mathcal{P}_j}{\partial \delta}\right)^2
\end{equation}

\noindent where $\mathcal{P}_j$ is probability associated with the successful projection in the corresponding mode $\ket{j}$. \\ 

\noindent In the small separation regime, where errors from direct imaging start to diverge, one can find an optimal basis of projective measurements for which the Fisher information remains constant. For a general point spread function(PSF) $\ket{\Psi(x)}$, the ``optimal'' mode of projection is proportional to its derivative which can be seen as follows,\\

\noindent The displaced PSFs $\braket{x_s}{\Psi}^{\pm}=\braket{x_s}{\Psi(x\pm s)}$, where $\delta=2s$, can be expanded in terms of the derivatives of the original PSF as,

\begin{equation} \label{eq:expansionSM}
\braket{x_s}{\Psi}^{\pm} = \sum_{n}\frac{(\pm s)^n\braket{x_s}{\Psi}^{(n)}}{n!}
\end{equation}

\noindent where $\braket{x_s}{\Psi}^{(n)}$ is the $n^{th}$ derivative in $x_s$ of $\braket{x_s}{\Psi}$. For small separations, we can consider only upto the first order term in the separation,

\begin{equation} \label{eq:approxphiSM}
\braket{x_s}{\Psi}^{\pm} \approx \braket{x_s}{\Psi} \pm s \braket{x_s}{\Psi}^{(1)} + O(s^2)
\end{equation}
\\
If the two displaced PSFs are incoherent then the resulting state is characterized by a completely mixed state,

\begin{equation} \label{eq:displacedDM}
\rho = \frac{1}{2}\left[\ket{\Psi}^{+}\!\bra{\Psi}+\ket{\Psi}^{-}\!\bra{\Psi}\right]
\end{equation}
where $\ket{\Psi}^{\pm}$ refer to the wavefunctions displaced by $\pm s$ in the $x$ axis.
\\
\noindent If we plug \ref{eq:approxphiSM} into \ref{eq:displacedDM} we find the density matrix approximated for small separations,

\begin{equation} \label{eq:approxDM}
\rho_ \approx \ket{\Psi}\bra{\Psi}+s^2\ket{\Psi}^{(1)}\bra{\Psi}^{(1)}+\frac{s^2}{2}\big ( \ket{\Psi}^{(2)}\bra{\Psi}+\ket{\Psi}\bra{\Psi}^{(2)}\big )
\end{equation}
\\
\noindent The probability of successful projection onto the normalized first derivative mode is,

\begin{flalign}
\mathcal{P}_{der} & \approx  \frac{1}{\mathcal{N}}\bra{\Psi}^{(1)}\rho_{BI}\ket{\Psi}^{(1)} \nonumber \\
& \approx \frac{s^2}{\mathcal{N}}|\bra{\Psi}^{(1)}\ket{\Psi}^{(1)}|^2 \nonumber \\
& \approx \frac{s^2}{\mathcal{N}} = \frac{\delta^2}{4\mathcal{N}},
\end{flalign}
\\
where $\mathcal{N}$ is the normalization factor. The Fisher information can be calculated as,
\begin{equation} \label{eq:fisheroptimalSM}
\mathcal{FI}_{der} = \frac{1}{\mathcal{P}_{der}}\left(\frac{\partial \mathcal{P}_{der}}{\partial \delta}\right)^2 = \mathcal{N}
\end{equation}

\noindent We see that in the small separation regime, the Fisher Information in the derivative mode is a constant regardless of the separation.

{\section{Schmidt Decomposition of SPDC}\label{SM: schmidt decomposition}}
\noindent
The bi-photon state generated in Spontaneous Parametric Down Conversion (SPDC) process can be decomposed into a superposition of tensor product of two linear combinations of Hermite-Gauss modes \ref{eq:tensorSM},

\begin{equation} \label{eq:tensorSM}
\ket{\Psi} = \sum_{m,n}\sum_{m',n'}C_{mn}^{m'n'}\ket{m,n}_s\otimes\ket{m',n'
}_i
\end{equation}

\noindent Hermite-Gauss $\mathcal{HG}$ modes $\ket{m,n}=\ket{m}\otimes \ket{n}$ form an orthonormal basis of spatial modes for the transverse plane and have the following expressions (2D and 1D respectively) in adimensional coordinates:

\begin{equation} \label{eq:HG2dSM}
\braket{x_o,y_o}{m,n} = \frac{1}{\sqrt{2^m2^n m!n! \pi}} \mathcal{H}_m(x_o) e^{-\frac{x^2_o}{2}} \otimes \mathcal{H}_n(y_o) e^{-\frac{y^2_o}{2}}
\end{equation}

\begin{equation} \label{eq:HD1dSM}
\braket{x_o}{m} = \frac{1}{\sqrt{2^m m! \sqrt{\pi}}} \mathcal{H}_m(x_o) e^{-\frac{x^2_o}{2}}
\end{equation}

\noindent where $\mathcal{H}_i(x_k)$ is the Hermite polynomial of the $i^{th}$ order for the variable $x_k$, and $x_o = \frac{\sqrt{2}}{\sigma_s}x$, $y_o=\frac{\sqrt{2}}{\sigma_s}y$ as defined in the main text. When expressed in the Schmidt basis of Hermite Gauss $\mathcal{HG}$ modes the original SPDC state can be simplified to \cite{miatto2012cartesian,straupe2011angular,SpatialSchmidt},

\begin{equation} \label{eq:schmidtSM}
\ket{\Psi} = \sum_{m,n}C_{mn}\ket{m,n}_s\otimes\ket{m,n}_i
\end{equation}

\noindent where $C_{mn} = \frac{4\gamma}{(1+\gamma)^2}|\frac{1-\gamma}{1+\gamma}|^{(m+n)}$ and where $\gamma$ is a constant determined by source parameters such as crystal length and pump beam waist. When performing any sort of joint measurement (i-e coincidence counts or coincidence imaging) on the entangled state, we consider the PSF of the form of the biphoton expression given in \ref{eq:schmidtSM}. If a separation is introduced in the signal beam, it becomes a mixed state of two separated beams with the signal photon displaced in the $x_s$ axis. The two-photon density matrix $\rho_{BI}$ resembles equation \ref{eq:displacedDM}, with $ \braket{x_s}{m^{\pm}}_s = \sqrt{ \frac{1}{2^{m} m! \sqrt{\pi}}} \mathcal{H}_m(x_s\pm s) e^{-\frac{(x_s\pm s)^2}{2}}$:

\begin{equation} \label{densitySPDCSM}
\rho_{BI} = \sum_{m,n}\sum_{u,v}C_{mn}C_{uv}^*\ket{m,n}_i\prescript{}{i}{\bra{u,v}}\otimes\bigg[\ket{m^{+},n}_s\prescript{}{s}{\bra{u^{+},v}} + \ket{m^{-},n}_s\prescript{}{s}{\bra{u^{-},v}}\bigg]
\end{equation}

\noindent The derivative of the PSF (\ref{eq:schmidtSM}) in $x_s$ has the following form:

$$\ket{\Psi}^{(1)}=\sum_{m,n}C_{mn}\ket{m}_s^{(1)}\ket{n}_s\otimes\ket{m,n}_i$$

\noindent where we can find from the recurrence relations of the Hermite Polynomials, that:

\begin{equation} \label{eq:hermitederivativeSM}
\ket{m}_s^{(1)}=\sqrt{\frac{m}{2}}\ket{m-1}_s-\sqrt{\frac{m+1}{2}}\ket{m+1}_s
\end{equation}

\noindent So, the derivative becomes a maximally entangled state with the expression:

\begin{equation}
\ket{\Psi}^{(1)}=\sum_{m,n}C_{m,n}\left(\sqrt{\frac{m}{2}}\ket{m-1}_s-\sqrt{\frac{m+1}{2}}\ket{m+1}_s\right)\ket{n}_s\otimes\ket{m,n}_i
\end{equation}

\noindent Projecting onto such a state is not trivial and is outside the scope of this experiment. We instead chose to use measurements readily available to us experimentally in a setup with entangled photon pairs. A variety of measurements (either categorized as joint or single-photon measurements) are possible which will determine the projection outcomes $\mathcal{P}_{i}$. Using a joint basis of $\mathcal{HG}$ modes as the mode sorting basis, with elements $\ket{k,l}_s\bra{k,l}_s \otimes \ket{k',l'}_i\bra{k',l'}_i$, the projection outcomes $\mathcal{P}_{kl}^{k'l'}$ can be expressed as:

\begin{flalign} 
\mathcal{P}_{kl}^{k',l'} & = \Tr[\ket{k,l}_s\bra{k,l}_s \otimes \ket{k',l'}_i\bra{k',l'}_i\rho_{BI}] \nonumber \\
& =  \bra{k,l}_s \otimes\bra{k',l'}_i\rho_{BI}\ket{k,l}_s\otimes\ket{k',l'}_i \nonumber \\
& = \frac{1}{2}C_{k',l}^2 \Big\{ |\langle{k}\ket{k'^+}|^2+|\langle{k}\ket{k'^-}|^2 \Big\}\delta_{l,l'} \label{eq:probexpressionSM}
\end{flalign} \\

\section{Analytical Expansion of $\braket{m}{n^{\pm}}$}\label{SM: overlap calculation}

\noindent The inner product between an $\mathcal{HG}$ function of mode m and a displaced $\mathcal{HG}$ function of mode n can be written out as:

$$\braket{m}{n^{\pm}} =\int_{-\infty}^{+\infty}\mathcal{HG}_m(x)\mathcal{HG}_n(x\pm s)dx$$

\noindent Applying a change of variables $u = x\pm \frac{s}{2}$ and inserting the expressions in \ref{eq:HD1dSM}:

\begin{flalign} \label{eq:innerproductSM}
\braket{m}{n^{\pm}} & = \frac{1}{\sqrt{2^m2^nm!n!\pi}}\int_{-\infty}^{+\infty}e^{-\frac{1}{2}[(u\mp \frac{s}{2})^2+(u\pm \frac{s}{2})^2]}\mathcal{H}_m(u\mp\frac{s}{2})\mathcal{H}_n(u\pm\frac{s}{2})du \nonumber \\
& = \frac{e^{-\frac{s^2}{4}}}{\sqrt{2^{m+n}m!n!\pi}}\int_{-\infty}^{+\infty}e^{-x^2}\mathcal{H}_m(u\mp\frac{s}{2})\mathcal{H}_n(u\pm\frac{s}{2})du 
\end{flalign}

\noindent The integral in \ref{eq:innerproductSM} is given in a table of integrals \cite{gradshteyn2014table}, where for $n \geq m$:
\begin{equation} \label{eq:hermiteintegralSM}
\int_{-\infty}^{+\infty}e^{-x^2}\mathcal{H}_m(x+y)\mathcal{H}_n(x+z)dx=2^n\pi m!z^{n-m}\mathcal{L}_m^{n-m}(-2yz)
\end{equation}

\noindent where $\mathcal{L}_i^{\alpha}$ is the generalized Laguerre polynomial of order i,$\alpha$. So, the analytical expression for the inner product is found to be, for $n \geq m$:

\begin{equation} \label{eq:analyticalSM}
\braket{m}{n^\pm} = \sqrt{\frac{m!}{n!}}2^{\frac{m-n}{2}}(\pm s)^{n-m}e^{-\frac{s^2}{4}}\mathcal{L}_m^{n-m}(\frac{s^2}{2})
\end{equation}

\section{Biphoton SPADE based on projection onto HG-modes}\label{SM:HG_biphotonSPADE}

\noindent Here, we derive the main result showing the relation between the Fisher information and the Schmidt rank ($K$) in biphoton SPADE. In the small separation limit (using \ref{eq:approxphiSM}, \ref{eq:hermitederivativeSM}) only three terms in \ref{eq:probexpressionSM} are non-zero:

$$
\mathcal{P}_{kl}^{k'l'} \approx C_{k',l}^2 \Bigg[ \delta_{k,k'}^2\big \{1-\frac{s^2}{2}(2k'+1)\big \} + \delta_{k,k'-1}^2\frac{k'}{2} +\delta_{k,k'+1}^2\frac{k'+1}{2} \Bigg]\delta_{l,l'}^2.
$$

\begin{equation} \label{eq:probapproxSM}
    \begin{cases}
      \mathcal{P}_{kl}^{kl} \approx C_{k,l}^2 (1-\frac{s^2}{2}(2k+1))\\ \\
      \mathcal{P}_{kl}^{k+1,l} \approx s^2C_{k+1,l}^2\frac{k+1}{2}\\ \\
      \mathcal{P}_{kl}^{k-1,l} \approx s^2C_{k-1,l}^2\frac{k}{2}
    \end{cases}       
\end{equation}
\\

\noindent For $k=k'$ the $\mathcal{FI}$ about the parameter $\delta = 2s$ tends to 0. For $k=k'\pm 1$ the projection outcome has the same form as for optimal measurements, where $\mathcal{P}_i$ is proportional to $\delta^2$.

\begin{equation} \label{eq:FIapproxSM}
    \begin{cases}
      \mathcal{FI}_{kl}^{kl}(s) \approx \frac{\big[C_{k,l}^2(2k+1)s\big]^2}{4-2s^2(2k+1)} \rightarrow 0\\ \\
      \mathcal{FI}_{kl}^{k+1,l} \approx \frac{k+1}{2}C_{k+1,l}^2\\ \\
      \mathcal{FI}_{kl}^{k-1,l} \approx \frac{k}{2}C_{k-1,l}^2
    \end{cases}       
\end{equation}
\\

\noindent For example, projecting signal photon onto $\ket{1,0}_s$ and coupling idler photon into a single mode fiber (projection onto $\ket{0,0}_i$) would give a constant Fisher Information, for small separations, of $\frac{1}{2}C_{0,0}^2$. The total $\mathcal{FI}$ is obtained by summing the non-vanishing $\mathcal{FI}$ across all $HG$ mode measurements. Writing out the coefficients explicitly:

\begin{equation*}
    \begin{cases}
      \mathcal{FI}_{kl}^{k+1,l} \approx \frac{k+1}{2}C_{0,0}^2\Big( \frac{1-\gamma}{1+\gamma} \Big)^{2(k+1)}\Big( \frac{1-\gamma}{1+\gamma} \Big)^{2l}\\ \\
      \mathcal{FI}_{kl}^{k-1,l} \approx \frac{1}{2}kC_{0,0}^2\Big( \frac{1-\gamma}{1+\gamma} \Big)^{2(k-1)}\Big( \frac{1-\gamma}{1+\gamma} \Big)^{2l}
    \end{cases}       
\end{equation*}
\\

\noindent and using $\sum_{k}k\Big( \frac{1-\gamma}{1+\gamma} \Big)^{2k} = \frac{(1-\gamma)^2(1+\gamma)^2}{16\gamma^2}$ , $\sum_{k}\Big( \frac{1-\gamma}{1+\gamma} \Big)^{2k} = \frac{(1+\gamma)^2}{4\gamma}$  we can calculate the sum over k for the 1D case \ref{eq:1dfiSM} or the sum over all $k,l$ \ref{eq:2dfiSM}.

\begin{equation} \label{eq:1dfiSM}
    \begin{cases}
      \sum_{k}\mathcal{FI}_{kl}^{k+1,l} = \frac{1}{2} (\frac{1-\gamma}{1+\gamma})^2 \approx 0.27\\ \\
      \sum_{k}\mathcal{FI}_{kl}^{k-1,l} = \frac{1}{2}
    \end{cases}       
\end{equation}

\begin{equation} \label{eq:2dfiSM}
    \begin{cases}
      \sum_{k,l}\mathcal{FI}_{kl}^{k+1,l} = \frac{(1-\gamma)^2}{8\gamma} \approx 0.60\\ \\
      \sum_{k,l}\mathcal{FI}_{kl}^{k-1,l} = \frac{(1+\gamma)^2}{8\gamma}\approx 1.10
    \end{cases}       
\end{equation}
\\

\noindent The numerical values correspond to the FI value for a $\gamma =0.15$, as in the experiment. Note that for strong spatial correlations, the value of $\gamma$ gets closer to 0 and the total FI increases. In fact, the schmidt number $K=\frac{(\gamma+\gamma^{-1})^2}{4}$, which is often used as a measure to describe the strength of spatial entanglement in the SPDC pair, can be found in the total FI sum: 

\begin{equation} \label{eq:fischmidtSM}
      \sum_{k,l}\mathcal{FI}_{coinc} = \frac{1}{4}(\gamma +\gamma^{-1}) = \frac{1}{2}\sqrt{K} \geq \frac{1}{2}
\end{equation}

\section{Gaussian Case ($\gamma = 1$)\label{SM:gaussian case}}
\noindent Note that $\gamma = 1$, corresponds to a simple separable product state of a two gaussian wavefunctions for signal and idler. In that case, the total Fisher Information according to Eq. \ref{eq:fischmidtSM} reduces to $\frac{1}{2} (\frac{N}{\sigma_s^2})$, which can be briefly seen as follows:

\noindent  The state after displacement can be represented by a density matrix $\rho = \frac{1}{2}\big[ \ket{0^+0}\bra{0^+0}+\ket{0^-0}\bra{0^-0}\big]$. 
At the small separation regime the optimal mode of projection is the first order Hermite Gauss mode $\mathcal{HG}_{01}$, for which the probability for a successful projection is given by,

$$P_1 = \bra{10}\rho\ket{10} = \frac{1}{2}\big[ |\braket{1}{0^+}|^2+|\braket{1}{0^-}|^2\big]$$
\\

\noindent Using equation \ref{eq:analyticalSM}, we have that $\braket{1}{0^+} = \braket{1^-}{0} = -\sqrt{\frac{1}{2}}s e^{-\frac{s^2}{4}}$ and that
$\braket{1}{0^-} = \braket{1^+}{0} = \sqrt{\frac{1}{2}}s e^{-\frac{s^2}{4}}$.
The probability outcome $P_1$ is given by:

$$P_1 = \frac{1}{2}s^2e^{-\frac{s^2}{2}}$$ 

\noindent with Fisher Information of:

$$FI_1 = \frac{1}{P_1}\left( \frac{\partial P_1}{\partial (2s)}\right ) =  \frac{1}{4}(2-2s^2+\frac{s^4}{2})e^{-\frac{s^2}{2}}\approx \frac{1}{2} (\frac{N}{\sigma_s^2})$$
\\

\noindent The total FI for the simple gaussian case does indeed correspond to the total FI when $\gamma = 1$. The entanglement scheme thus provides an advantage over classical SPADE for any level of entanglement with $\gamma<1$.

%%%%%%%%%%%%%%%%%%%%%%%%%%%%%%FIGURES%%%%%%%%%%%%%%%%%%%%%%%%%%%%%%%%%%%%%%%%%%%%%%%%%%%%%%%%%%%%%%%%%%%%%%%%%%%%%%%%%%%%%%%%%%%%%%%%%%%%%%%%%%%%%%%%%%%%%%%%%%%%%%%%%%%%%%%%%%%%%%%%%%
\end{document}